\documentstyle[psfig,multicol,aps,epsf]{revtex}
\begin{document}
\twocolumn[\hsize\textwidth\columnwidth\hsize\csname
 @twocolumnfalse\endcsname
\title{Rain: Relaxations in the sky}
\author{Ole Peters$^{*}$ and Kim 
Christensen} 
\pagestyle{myheadings} 
\address{Blackett Laboratory, Imperial College, Prince Consort 
Road, 
London SW7 2BW, United Kingdom} 

\maketitle 
\begin{abstract} 
We demonstrate how, from the point of view of energy flow through an open system, rain is analogous to many other 
relaxational processes in Nature such as earthquakes. By identifying rain events as the basic entities of the phenomenon, 
we show that the number density of rain events per year is inversely proportional to the released water column raised to 
the power 1.4. This is the rain-equivalent of the Gutenberg-Richter law for earthquakes. The event durations and the 
waiting times between events are also characterised by scaling regions, where no typical time scale exists. The Hurst 
exponent of the rain intensity signal $H = 0.76 > 0.5$. It is valid in the temporal range from minutes up to the full 
duration of the signal of half a year. All of our findings are consistent with the concept of self-organised criticality, 
which refers to the tendency of slowly driven non-equilibrium systems towards a state of scale free behaviour.
\end{abstract} 
\vskip2pc] 

\section{Introduction} 
\label{introduction} 
Water is a precondition for human survival and civilisation. For this 
reason, measurements on water resources have been recorded for several 
centuries. A time series from the Roda gauge at the Nile reaches back to 
the year 622 AD \cite{1}. The main focus of analysis has historically 
been on statistics yielding a reliable estimate for the rainfall during 
the growth season. The most obvious question to ask is in this context: How 
much does it rain, on average, in the relevant months? Questions of this 
type can be answered using long time series without high temporal 
resolution, and a measurement of relatively low sensitivity may be 
sufficient. Entirely different levels of resolution and precision are 
needed in order to penetrate further into the complexity of 
precipitation processes. One might want to know just how reliable -- or 
in fact how meaningful -- is an estimate of future rainfall based on 
averages from the past. Of course, one would ultimately like to 
understand the processes that make a cloud release its water. Questions 
of this kind point to the statistical properties of rain events rather 
than temporal averages. 

In Sec. \ref{measurement} we discuss the new type of radar measurement on which our analyses are based. A time-series of 
high precision rain rates with one minute resolution was obtained. Section \ref{dataanalysis} is subsectioned and 
introduces the various measures we apply to the time series. In Sec. \ref{eventsizes} we introduce the fundamental concept 
of rain events as sequences of non-zero rain rates, which enables comparison with many other relaxational processes 
endowed 
with an event-like structure 
\cite{2}. Equipped with this concept we investigate the statistical properties of event sizes. Over at least 
three orders of magnitude of event sizes the number density is consistent with a decaying power law implying 
that there is no typical event size. We find that the most frequent small events are considerably below the typical 
sensitivity threshold of standard rain gauges \cite{3}. 
In Sec. \ref{eventduration} and \ref{droughtduration} we consider the event durations and the waiting times between 
successive events. The number densities of both event durations and waiting times follow power laws. In addition, we note 
a non-trivial relation between the duration and the size of events.
In Sec. \ref{fractaldimension} we define the binary signal in time of either rain or no rain and relate 
the probability distribution of waiting times to the fractal dimension of this signal. It is 
then speculated that the physical reason for the lower breakdown of the observed fractal regime at a time-scale of the 
order of $10$ min may be set by the time it takes for cloud droplets to grow into raindrops. The upper end of the scaling 
region coincides with the time scale given by passing frontal weather systems. In Sec.  \ref{hurst} we determine the Hurst 
exponent of the rain signal as 0.76, spanning four orders of magnitude
$\tau \hspace{0.2cm} \epsilon \hspace{0.2cm}[10 \; \mbox{min}, 1/2 \; \mbox{year}]$, extending Hurst's result from the 
Nile gauge at Roda, which is valid for $\tau \hspace{0.2cm} \epsilon \hspace{0.2cm} [1 \; \mbox{year}, 1080 \; 
\mbox{years}]$. Section \ref{context} establishes a close analogy between the observed characteristics and other 
relaxational processes such as earthquakes and avalanches in granular media. Finally in Sec.  \ref{conclusion} we conclude 
that the framework of self-organised criticality may serve as a useful working paradigm when dealing with rain. 
\section{Measurement} 
\label{measurement} 
The recent developments in remote sensing techniques have opened entirely new opportunities to rain analysis. By using a 
radar rather than a common water gathering device, the limits on rain measurements due to evaporation, sensitivity 
threshold, averaging times and accessibility can be pushed considerably \cite{4}. 

The data we used refer to a height range of $50$ m at $250$ m above sea level and have been collected from January to July 
1999 with the Micro Rain Radar MRR-2, developed by METEK \cite{5}. The radar is operated by the Max-Planck-Institute 
for Meteorology, Hamburg, in Germany at the Baltic coast in Zingst (54.43$^{\circ}$N 12.67$^{\circ}$E) under the 
Precipitation and Evaporation Project (PEP) in BALTEX \cite{6}. The retrieval of the rain rate is based on a Doppler  
spectrum analysis described by Atlas \cite{7}. At vertical incidence, the fall velocity of a droplet can be identified 
with 
the Doppler shift. The friction  force acting on a falling drop increases approximately proportional to its surface, but 
the gravitational force increases proportionally to its volume. Therefore, in the atmosphere, larger drops fall faster 
than 
smaller ones, and spectral bins can be attributed to corresponding drop sizes. For a given drop size, scattering cross 
sections can be calculated by Mie theory \cite{8}. Droplets are approximated by ellipsoids with known axis ratio \cite{9}. 
The influence of the changing air density with height is considered according to Beard \cite{10}, and standard atmospheric 
conditions are assumed \cite{11}. Attenuation of radar waves by droplets is accounted for using the observed droplet 
spectrum of the lowest range gate to estimate attenuation for the following one. For higher gates all observed and 
corrected spectra of lower layers are taken into account. Thus, from the Doppler spectrum alone one can infer the number 
of 
drops $n_i$ of any desired volume $V_i$ as well as their fall velocities $v_i$. The rain rate can be calculated 
instantaneously as $q(t)=\sum_i n_i V_i v_i$. In the time series we investigated, the continuous measurement is averaged 
over one-minute intervals, leading to one minute temporal resolution. When the signal due to rain becomes 
indistinguishable 
from the background noise at the receiver, the rain rate is defined as zero. Under the pertinent conditions, the 
calculated 
rain rate was typically $q_{min}=0.005$ mm/h, when this happened. What is measured at this sensitivity threshold would 
probably more sensibly be labelled the turbulent motion of drizzle through the atmosphere, rather than rain. Instead of 
asking whether rain can be detected, a question that arises now is what we actually mean by rain. To achieve this level of 
precision, a conventional pluviometer would have to be able to detect a water column of $83.3$ nm ``rain'' spread out over 
one minute. For comparison, the diameter of a single water molecule is about $0.3$ nm. Thinking in terms of accumulated 
water column in such a rain gauge and given one-minute averaging time, one would come to the conclusion that the smallest 
detectable rain event corresponds to a minute during which on average every second, a film four molecules thick, drifts 
down towards the ground. This, of course, would be impossible to detect. The MRR-2, however, employs a method that is not 
based on water hitting or passing through an area of a few decimeters across. Given the $50$ m height range starting at 
$250$ m above the radar with $2^{\circ}$ opening 
angle, it measures what happens to the liquid water in a volume of the order of $1000$ m$^{3}$, and one must bear that in 
mind to understand the minimum values calculated above. Of course, events at the radar's sensitivity threshold are far 
from 
being detectable by any water-collecting pluviometer and similarly far away from what we associate with the word ``rain''. 
Nonetheless, we will consider any minute with derived $q(t) > q_{min}$ as ``rain'', and conversely, only if the radar 
fails 
to detect any net downward motion of water through the air, we will speak of ``no rain''. We will come back to this point 
in Sec. \ref{eventsizes}. Especially for small rain rates the employed method is extremely powerful. 

The quantitative retrieval is restricted to rain. The reflection
spectra of snow and hail look very different from those of 
liquid water and can be identified. In this case the method
fails to calculate correct water masses. The latest 
version of the instrument recognises non-rain precipitation
by an internal algorithm. The rain intensity data used were 
calculated from measurements performed while the development
of the instrument was still ongoing, and hence the raw data 
had to be checked manually. 
\section{Data Analysis} 
\label{dataanalysis} 
\begin{figure}[h] 
\vspace*{-1.5cm} 
\centerline{\psfig{figure=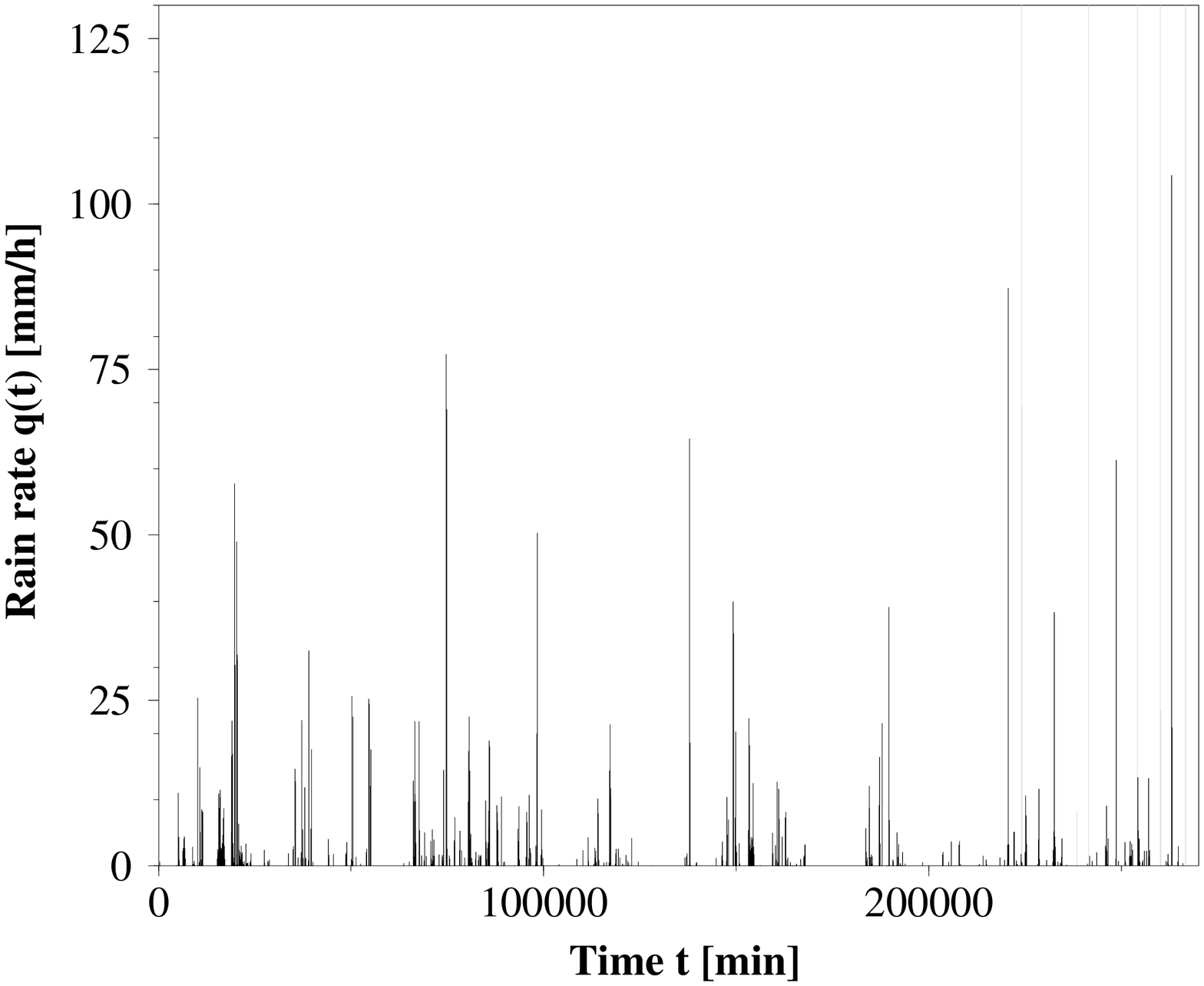,height=7cm,angle=-90}} 
\vspace*{-0.05cm} 
\caption{The rain rate [mm/h], averaged over one minute, plotted versus the time of occurrence [min since 01-01-99, 0:00].  
The five high peaks at the right-hand side of the figure, shown as dashed lines, are a result of malfunctioning during 
extreme weather conditions in thunderstorms. They correspond to $75$ out of $266,611$ minutes, which is so small a 
fraction 
that none of our results would be significantly altered by including them.} 
\label{fig1} 
\end{figure}

The months January and February contain several instances of snow at our chosen measuring height of $250$ m. By far the 
largest snow disturbance was observed on March 6, from 3:49 am until 11:38 pm. The Doppler spectra reveal that 
$250$ m altitude was inside the melting layer, and the water column resulting from interpreting the event as rain would 
have been $279$ mm, which is of the order of the usual rainfall of eight weeks. In June and July, five very short periods 
of extremely high calculated rain rates were found (see Fig. \ref{fig1}). The Doppler spectra indicate two different 
types of drops with fall velocities at $\approx4$ m/s and $\approx9$ m/s. Comparison with meteorologic records shows that 
around these times, thunderstorms with hail or extreme rainfall may have caused the radar to malfunction. As in the case 
of 
snow disturbances, data gathered during these periods were excluded from the analyses in Sec. \ref{eventsizes} and 
\ref{hurst}. The results in Sec. \ref{eventduration}, \ref{droughtduration}, and \ref{fractaldimension}, however, refer to 
the entire data set since the value of the rain intensity is irrelevant here. To make sure that our results are not an 
artifact of the observed anomalies, all analyses were also performed on the certainly clean months of April and May. No 
differences to the previously obtained results were observed. Due to the high resolution, not even the ranges of validity 
were significantly affected. 

\subsection{Event Sizes} 
\label{eventsizes} 
Previous work focused on rainfall during fixed time intervals and on the statistical properties of such fluctuating rain 
intensities. Other studies addressed distributions of wet and dry spells (see e.g. \cite{12}). The fundamental novelty of 
the present study is to acknowledge the event-like structure of rain \cite{2}. Events are defined as a sequence 
of non-zero rain rates, and their size $M=\sum_{t} q(t) \Delta t$, with $\Delta t = 1$ min, is the accumulated water 
column 
during the event. The intervals 
of zero rain rate between events are called drought periods. Our perspective is motivated by work on other Natural 
phenomena, such as earthquakes, where one is mainly interested in the events. While the entire agricultural sector depends 
on a sufficient amount of rain spread out over the months of the growth season, no one depends on the average seasonal 
flow 
of energy through the earth's crust. Due to this difference in anthropogenic interest, the two perspectives have been used 
almost entirely separately in the respective fields. Owing to the precision and high temporal resolution of the data, an 
investigation into the fine structure of rain events was made possible, and the results are strikingly clear. 
\begin{figure}[h] 
\vspace*{-1.5cm} 
\centerline{\psfig{figure=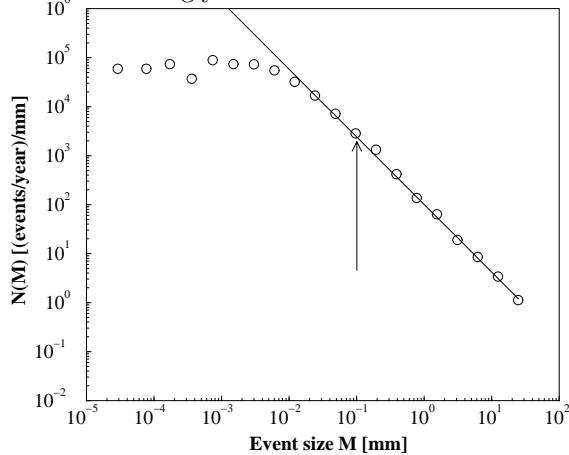,height=7cm,angle=-90}} 
\vspace*{0.1cm} 
\caption{The number density N(M) of rain events versus the event size $M$ (open circles) on a double logarithmic scale. 
Events are collected in bins of exponentially increasing widths. The horizontal position of a data point corresponds to 
the 
geometric mean of the beginning and the end of a bin. The vertical position is the number of events in that bin 
divided by the bin size. To facilitate comparison with future work, we rescaled the number of events to annual values by 
dividing by the fraction of a whole year during which the data were collected. The experimental data are consistent with a 
power law $N(M) \propto M^{-\tau_{M}}, \tau_{M} \approx 1.4$ (solid line) over at least three orders of magnitude, 
$M\hspace{0.2cm}\epsilon\hspace{0.2cm} [M_{min}, M_{max}]$ with $M_{min}\approx5*10^{-3}$ mm and $M_{max}\approx 35$ mm. 
The arrow indicates the typical sensitivity threshold of a conventional high precision tipping bucket rain gauge. Not only 
can we see that the radar technique is roughly 10,000 times more precise but also that a considerable fraction of rain 
events must be missed with conventional methods.} 
\label{fig2} 
\end{figure}
Figure 
\ref{fig2} shows the number density of rain events per year N(M) versus event size M on a double logarithmic plot. In a 
scaling regime $M_{min} < M < M_{max}$ extending over at least three orders of magnitude, the distribution follows the 
simple power law 
\begin{equation} 
N(M) \propto  M^{-\tau_{M}}, \hspace*{1cm}\tau_{M} \approx 1.4 . 
\label{powermass} 
\end{equation} 
This implies that a typical scale of events does not exist, and scale invariance prevails. In the scaling region, if we 
compare the frequency of events of size $M$ to that of events of size $kM$ we obtain the same fraction, independent of M. 
From Eq. (\ref{powermass}), it follows that: 
\begin{equation} 
N(M)/N(k*M)=k^{\tau_{M}},  \hspace*{0.75cm}M \hspace{0.2cm}\epsilon \hspace{0.2cm}[M_{min}, M_{max}]. 
\label{scaling}
\end{equation} 

But Fig. \ref{fig2} contains even more information. For events smaller than $M_{min}\approx 5*10^{-3}$ mm the power law 
breaks down. This is indicative of a different physical process being responsible for events in this realm. Within the 
scaling regime, events of all sizes look alike when compared to others. Hence there is no reason to assume different 
physical origins. We will later motivate the suggestion that this common origin is sudden relaxation, bursts of 
intermediately stored energy leaving the atmosphere. Where the power law breaks down, a different type of process sets in. 
Events smaller than $M_{min}$ might be due chiefly to the inner dynamics of the atmosphere. Virga, drizzle that evaporates 
before reaching the ground, is difficult to interpret from the event perspective. Drizzle can form at the lower edge of 
clouds but immediately re-evaporate. Commonly the distinction between cloud droplets and rain drops is made in terms of 
diameter.
When the droplet diameter surpasses $0.1$ mm one speaks of rain drops. This definition reflects a physical separation 
apparent from a gap in the drop size distribution around $0.1$ mm diameter \cite{13}. Fringes of virga, half cloud and 
half 
rain, may be the explanation for events smaller than $M_{min}$ in Fig. \ref{fig2}. Indicated with an arrow in Fig. 
\ref{fig2} is the typical sensitivity threshold $0.1$ mm of high precision tipping bucket rain gauges. The value $0.1$ mm  
is widely used as the definition of zero precipitation \cite{3}. Given that our interpretation of the breakdown of the 
power law is correct, and every rain event with $M>5*10^{-3}$ mm is actual rain, it is evident that measurements with 
today's standard precision simply don't see a considerable fraction of the rain events. Questions regarding the fine 
structure of rain and the actual physical processes involved are then hard to address. With the radar 
measurement on the other hand, all rain seems to be captured and we can choose a suitable limit ($M_{min}$) below which 
events are ascribed to a different physical process. 

To ascertain that we are capturing the entire physically relevant range of the observables of the process of rain, it is 
evidently necessary to use observational techniques enabling us to see beyond the physical limits of rain. Results from 
investigations that do not fulfill this requirement cannot be conclusive and must be treated with careful scepticism. The 
present study suggests a reasonable maximum sensitivity threshold of around $5*10^{-3}$ mm, which is one twentieth of the 
commonly used threshold.

Assuming Eq. (\ref{powermass}), we can easily calculate the number $N(M>M_{1})$ of expected events exceeding a given mass 
$M_{1}$.

\begin{equation}
N(M>M_{1})\propto \int_{M_{1}}^{\infty}M^{-\tau_{M}} dM =\frac{1}{\tau_{M}-1}M_{1}^{-\tau_{M}+1}.
\end{equation}

It follows that 

\begin{equation}
{N(M>M_{2})=N(M>M_{1})\left(\frac{M_{2}}{M_{1}}\right)^{-\tau_{M}+1}} .
\label{ratio}
\end{equation}

Since we know how many events there are with $M>M_{1}=M_{min}$, Eq. (\ref{ratio}) can be used to estimate $N(M>M_{2})$, 
where 
$ M_{2}>M_{min}$. We observed 10 events in the largest non-empty bin ranging from $17$ mm to $35$ mm, but from 
extrapolating the power law as outlined above, we would another 10 in the following bin ranging from $35$ mm to $70$ mm. 
In 
total we would expect to see 38 events larger than the largest event that was actually observed. We therefore conclude 
that the sudden upper cutoff apparent in Fig. \ref{fig2} is not due to the limited time of observation but rather 
reflects a physical limit to the process of rain at the given location. We define $M_{max}$ as the largest event in the 
data set, a downpour of $M_{max}\equiv 35$ mm of rain.

\subsection{Event Duration} 
\label{eventduration} 
\begin{figure}[h] 
\vspace*{-1.5cm} 
\centerline{\psfig{figure=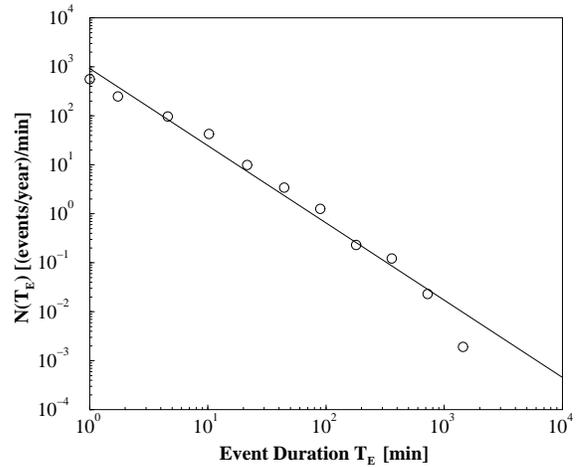,height=7cm,angle=-90}} 
\vspace*{0.1cm} 
\caption{The measured distribution of event durations (open circles). 
The data are consistent with a power law decay (solid line). The exponent of $\tau_{E}\approx 1.6$ is larger than the 
exponent for the 
event size distribution, implying a non-trivial relation between event duration and the rain rate during the event.} 
\label{fig3} 
\end{figure}

The number density of events versus event duration $T_{E}$ was found to approximate to a power law, see Fig. \ref{fig3}

\begin{equation}
N(T_{E})\propto T_{E}^{-\tau_{E}}, \hspace*{1cm} \tau_{E} \approx 1.6 .
\end{equation}
To highlight the implications of this result we consider the simplest form of a precipitation model. 
Naively, one might divide the number of minutes with measured rain by the total number of minutes observed, and simply use 
this fraction as the rain probability $p_{rain}$ in every minute. About 8\% of the minutes we observed contain rain. 
Therefore, the probability for two successive rain minutes would be $p_{rain}^{2} = 0.0064$, and for 5 successive minutes, 
it would already be negligible. Any model based on independent events produces characteristic time scales. In this case, 
one of the characteristic time scales would be the typical rain duration $t_r$. The probability for a rain event of 
duration $T_{E}$ is given by $p(T_{E})=p_{rain}^{T_{E}}$. This can be re-written as $p(T_{E})=e^{-T_{E}/t_{r}}$, where 
$t_{r}$ is the characteristic rain duration, which hardly any events will surpass. It follows that $t_{r}=-\frac{1} 
{ln(p_{rain})} \approx 24$ s. 

But the measured distribution is qualitatively different. Not only does the  power law like number density allow for 
events 
longer than $1000$ min, but no typical duration is found at all. We do not observe an exponential distribution of any 
kind.
 
The exponent in the power law relating the duration of events to their frequency is different from that for the event 
sizes. This implies a non-trivial relationship between the duration and the average rainrate during an event. If we could 
simply assume an average rain rate, equal for all rain events, the size would be proportional to the duration and the 
distributions would have the same exponent. Apparently, longer rain events are more intense.

\begin{figure}[h] 
\vspace*{-1.5cm} 
\centerline{\psfig{figure=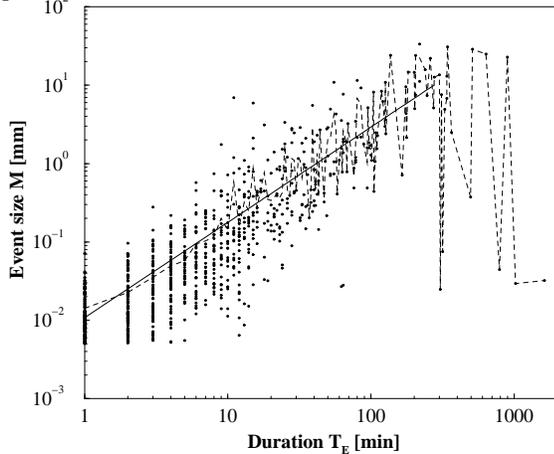,height=7cm,angle=-90}}
\vspace*{0.1cm} 
\caption{The event size versus event duration. The dots 
represent single events of the corresponding duration. For each 
duration, the average of the single dots is evaluated (dashed line). Up to about $200$ min the 
average event size $\langle M \rangle$(t) increases to a good 
approximation like $T^{1.2}$ (solid line). An exponent 
greater than 1 is consistent with Figs. \ref{fig2} and \ref{fig3}. 
Hence, on average the rain rate was greater for long events.} 
\label{fig4} 
\end{figure}

The statistical support for a difference between the exponents of event size and duration is not very strong but the 
results shown in Fig. \ref{fig4} reinforce this conjecture. Figure \ref{fig4} shows the average event size 
plotted versus event duration, and an exponent slightly greater than 1 is observed. This is a crude and somewhat forced 
measure to apply but it yields results that are qualitatively consistent with Figs. \ref{fig2} and \ref{fig3}, since if 
the average event size increased proportionally to the duration, the observed exponent would be 1. 

\subsection{Drought Duration} 
\label{droughtduration} 
\begin{figure}[h] 
\vspace*{-1.5cm} 
\centerline{\psfig{figure=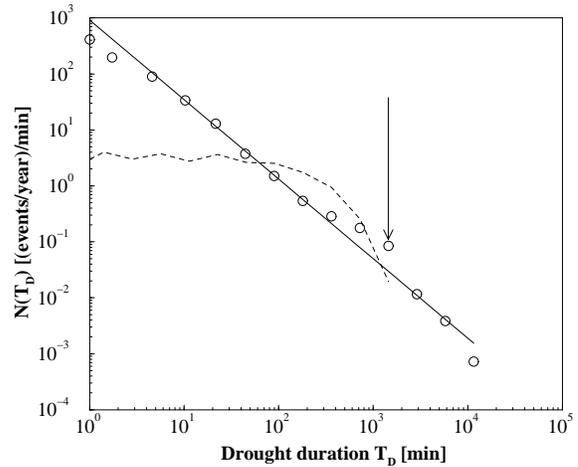,height=7cm,angle=-90}} 
\vspace*{0.1cm} 
\caption{The open circles show the number density $N(T_D)$ of drought periods per year versus the drought duration $T_D$. 
The solid line represents a power law approximation, with exponent $\tau_{D}=1.4$, to the observed distribution. The arrow 
indicates one day, around which 
a deviation form pure power law behaviour can be observed. This is due to the daily meteorological cycle. For comparison, 
a 
Poisson process, yielding an exponential distribution of waiting times, was fitted to the data (dashed line). The 
rate of events $\lambda$ is defined as the total number of observed rain events $N_{total}$ divided by the total time of 
observation $t_{total}$. The number of events is then normalised to annual values. The Poisson process would give a number 
density $N(10000$ min$)=1.2*10^{-18}$ (not shown). Clearly, the observed values are incompatible with such an uncorrelated 
process.} 
\label{fig5} 
\end{figure}

In Fig. \ref{fig5}, the probability distribution of drought durations $N(T_{D})$ is shown to follow a power law: 

\begin{equation} 
N(T_{D}) \propto T_{D}^{-\tau_{D}}, \hspace{1cm} \tau_{D}\approx 1.4 . 
\label{powerdrought} 
\end{equation} 

No cut-offs were apparent. The power law is a good approximation from the minimum (1 minute) all the way to the maximum 
(two weeks) of observed drought durations. The only observed deviation at droughts of around one day length is due to the 
daily meteorological cycle. As for the event durations, this behaviour clearly implies correlation. We can define the 
drought probability as $p_{drought}=1-p_{rain}$. Hence all the arguments in sec. \ref{eventduration} apply for drought 
durations too. With $p_{drought}$ replacing $p_{rain}$, the typical drought duration $t_{d}=-\frac{1}{ln(p_{drought})} 
\approx 12$ min. The dashed line in Fig. \ref{fig5} was generated by another method. Instead of treating minutes as the 
independent entities, we determine the rate at which rain events start by dividing the number of minutes by the number of 
rain events. This treatment takes into account the clustering of zero 
rain rates on the time axis, i.e. the persistence of droughts, but it cannot pay tribute to the dependencies which produce 
the real power-law behaviour. If arithmetic rather than exponential behaviour persists for more than two weeks then rain 
rates at times $t_{1}$ and $t_{2}$ two weeks apart still cannot be treated as independent. 

Adding the persistence of rain to that of droughts, the signal $q(t)$ can be modelled with a two-state Markov process. One 
then defines transition probabilities from rain to drought and drought to rain, consistently with the fraction of total 
rain and drought times. In this case, typical drought and rain durations can be chosen. Persistence is now accounted for, 
but the probability for observing drought or event durations above these characteristic time scales would still decay 
exponentially, while remaining constant for shorter events and droughts (see Fig. \ref{fig5}). This is incompatible 
with the observed distributions for both the drought durations and the event durations. The following section will 
strengthen this result further.

\subsection{Fractal Dimension} 
\label{fractaldimension} 
A fractal (see e.g. \cite{14}) is a structure displaying scale invariance of the type mentioned in Sec. \ref{eventsizes}. 
Zooming into a fractal with a factor of b and then re-scaling the coordinate system with a factor of $b^{d_{f}}$, where 
$d_{f}$ is called the fractal dimension, leaves the structure unchanged. Fractals often occur naturally, in which case the 
unchanged property is usually a statistical one. The rain data are from one fixed location but they span a long period of 
time. We define a binary signal -- either rain or drought -- and determine its fractal dimension in time, using the box 
counting method: Different lengths $l$ of time intervals (boxes) are used to cover the rainy sections on the time axis. 
The number of boxes $n(l)$ needed to cover the rain is proportional to $l^{-d_{f}}$.

\begin{figure}[h] 
\vspace*{-1.5cm} 
\centerline{\psfig{figure=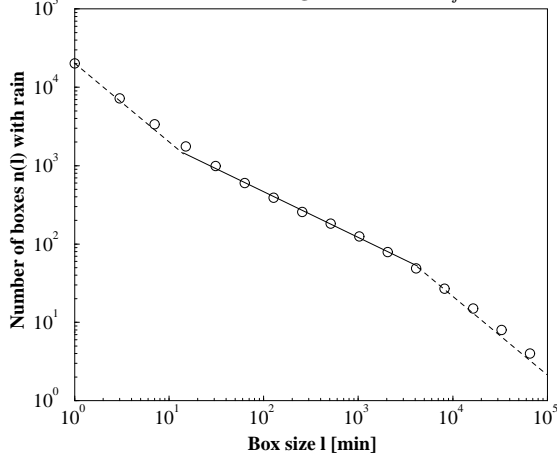,height=7cm,angle=-90}} 
\vspace*{0.1cm} 
\caption{The number of time intervals (boxes) needed to cover the rain versus the box size. The fractal dimension $d_{f}$ 
is minus the slope of this function in a double logarithmic plot. $d_{f} \approx 0.55$ in a scaling regime spanning $2$ 
orders of magnitude. Outside the scaling regime it assumes the trivial value $1$.} 
\label{fig6} 
\end{figure}

The results are displayed in Fig. \ref{fig6}. In the double-logarithmic plot we find an S-shaped curve. The dashed 
lines indicate two regimes with trivial slope, $d_{f}=1$, and the solid line a non-trivial regime where $d_{f}\approx0.55$. 
Consider again the simple model with the two state Markov chain. As long as the box size is below the typical rain 
duration, the number of boxes needed to cover the rain decreases trivially; they are used to fill the compact space of the 
rain events. When the typical rain duration is passed, each rain event is essentially covered by one box and the number of 
boxes remains constant. As the box size approaches the typical length for droughts, the entire duration of the measurement 
is filled, and the number of boxes begins to decrease in the trivial fashion again. The only way to obtain a non-trivial 
fractal dimension is to have -- in a sense -- typical droughts at all time scales. This amounts, of course, to having no 
typical drought duration at all. Mathematically, this scale-freedom is represented by the power-law distribution of 
drought 
durations. The number of boxes needed to cover the rain signal will be the true rain duration plus the time spanned by 
droughts that are shorter than the box size (these will be overlooked), all divided by the box size. Hence, apart from a 
constant, representing 8\% of the total time, the time $T_{c}$ spanned by the boxes to cover the rain will 
increase with $l$ as $ T_{c}=\int_{0}^{l} N(T_{D})*T_{D}dT_{D} \propto \int_{0}^{l} T_{D}^{-1.42}*T_{D}dT_{D}$, which is 
implied by Fig. \ref{fig5}. Evaluating the integral we have $T_{c}\propto l^{0.58} $. The number of boxes needed is 
$T_{c}/l=l^{-0.42}$. In this sense a fractal relation like the one shown in Fig. \ref{fig6} could be a  consequence 
of a power-law distribution of drought durations like in Fig. \ref{fig5}. It is the scale freedom that stretches the 
transition between the regime where the temporal resolution suffices to register the droughts and the regime where it does 
not. The values we measure suggest that there is more to the rain - no rain signal than only the power law of 
interoccurrence times. Deducing the fractal dimension from the drought distribution only, we would expect a value of 0.42. 
But we observe 0.55, and the difference appears to be significant. 

The scaling regime extends from a lower limit around $10$ minutes to an upper breakdown near $3$ to $4$ days. While one 
might expect the fractal regime to span further for longer time series, the analysis of a 30 year time-series from Uccle 
\cite{12} suggests that the observed breakdown is not an artifact of the shortness of our data-set. The authors place the 
cut-off at $3.5$ days, which coincides with our value. Apparently, the correlation that gave rise to the fractal relation 
does not hold for longer than $3.5$ days. Investigation of time series from Denmark with 1-day resolution, collected from 
1876 until 2000 \cite{15} suggests that the power law for droughts does not hold for drought durations exceeding the upper 
cut-off in the fractal dimension. 

The explanation for the upper cutoff of the fractal regime may be that the typical duration of a frontal system moving in 
from the Atlantic is of the order of 3 days. Measured rain parameters will not belong to the same frontal system if the 
measurements are temporally separated by significantly more than three days. The lower breakdown around $10$ min could not 
be observed in the Uccle time series since there the temporal resolution was only 10 minutes. We are still unsure as to 
how 
to interpret this lower breakdown. Clearly there must be a lower breakdown somewhere, and we expect it to occur where the 
particular kind of correlation that gave rise to the fractality on hourly to daily time scales ceases. The lower breakdown 
indicates that $10$ min is a time scale which is special, and it must be related to a physical process. The microphysical 
processes of coagulation that trigger a cloud to release its water content take place on this time scale. Starting with 
typical small cloud droplets with radius $r\approx 10^{-3}$ mm, the process of stochastic collection during which small 
droplets merge to form rain drops of appreciable fall velocity takes roughly $10-30$ min under typical warm cloud 
conditions \cite{16}. It is possible that coagulation starts at a certain level inside a cloud and then pauses at that 
level before a single drop has left the cloud. If it then starts again, it is possible that on the ground we observe two 
layers of rain separated by a vertical distance corresponding to up to $\approx 10$ min fall time. While this seems like 
two different events, from the cloud's perspective it is really only one, since the process of releasing water did not 
stop 
at any moment everywhere within the cloud. Effects of motion of the cloud relative to the ground are not included in these 
considerations. It is unlikely that the 10 minute time scale is a result of the employed measurement technique. The radar 
only picks up drops with appreciable fall velocity, $v>0.5$m/s. Thus the limit on the time resolution, given by the height 
of the scattering volume, is $50$ m/$0.5$ (m/s) $= 100$ s, for the slowest drops.

\subsection{Hurst Exponent} 
\label{hurst} 
In an attempt to determine the necessary size of a water reservoir that would never empty nor overflow, Hurst \cite{1} 
considered an incoming signal q(t), corresponding to the rain intensity in our case, that causes the level of a reservoir 
to rise or fall. Using our data, the deviation from the average water level in an imaginary reservoir would be 

\begin{equation} 
X(t,\tau)=\sum_{u=0}^{t}\left( q(t) - \langle q 
\rangle_{\tau}\right)\Delta t, 
\label{X} 
\end{equation} 
where $\Delta$ t = 1 min and 
\begin{equation} 
{\langle q \rangle}_{\tau} = \frac{1}{\tau} \sum_{t=1}^{\tau} q(t) .
\end{equation} 

\begin{figure}[h] 
\vspace*{-1.5cm} 
\centerline{\psfig{figure=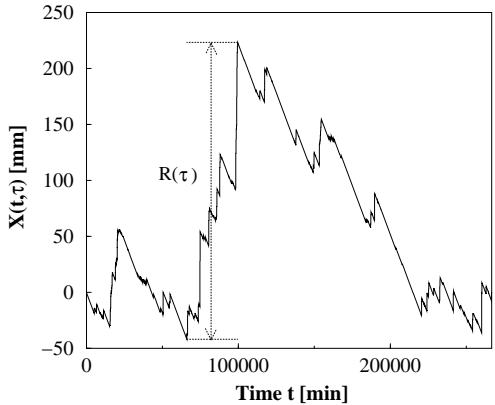,height=7cm,angle=-90}} 
\vspace*{0.1cm} 
\caption{Water level $X(t,\tau)$ in mm in an imaginary reservoir for $\tau$=266,611 min, as derived from Fig. 
\ref{fig1}. During drought periods, a constant, slow decrease in the water level is observed, whilst during rain events 
the water level increases rapidly. The necessary size of a sufficiently large reservoir is given by the range indicated by 
a dashed line.} 
\label{fig7} 
\end{figure}

\begin{figure}[h] 
\vspace*{-1.5cm} 
\centerline{\psfig{figure=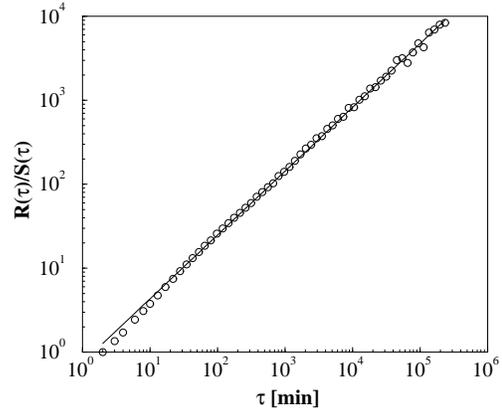,height=7cm,angle=-90}} 
\vspace*{0.1cm} 
\caption{The dimensionless ratio $R(\tau)/S(\tau)$ versus $\tau$ (open 
circles) shown on a double logarithmic scale. The slope of the fitted 
straight line (solid) reveals the anomalous Hurst exponent: 
$R(\tau)/S(\tau) \propto \tau^{H}$ with $H \approx 0.76$. The data 
deviate from the power law fit below $\tau \approx 10$ min in the lower 
limit, but no upper limit of the relation is observed.} 
\label{fig8} 
\end{figure}

The quantity -$\langle q(t) \rangle_{\tau}$ in Eq. (\ref{X}) can be thought of as an average outflux from the reservoir 
and 
insures that for any period $\tau$ the water level starts and ends at zero. Overall trends during the interval $\tau$ are 
thus eliminated. Figure \ref{fig7} shows $X(t,\tau)$ as derived from the data set in Fig. \ref{fig1}. 
The range of water levels the reservoir has to allow for is then given by 
\begin{equation} 
R(\tau) = \max_{1 \leq t \leq \tau} X(t,\tau) - \min_{1 \leq t \leq 
\tau} X(t,\tau). 
\end{equation} 
Hurst determined the dimensionless ratio $R(\tau)/S(\tau)$ as a 
function of $\tau$, where $S(\tau)$ is the standard deviation of the 
influx $q(t)$ in the period $\tau$. It can be shown that if $q(t)$ is 
any random signal with finite variance \cite{17}, this ratio increases 
as \begin{equation} R(\tau)/S(\tau) \propto {\tau}^H, \end{equation} 
where $H=1/2$ is called the Hurst exponent. Hurst's analysis on data 
from the Roda gauge at the Nile, however, yielded a different exponent of $H 
\approx 0.77$. 
This unexpected result is commonly interpreted as a sign of 
persistence in the signal, or even as correlation. 
The exponent obtained from performing the same analysis 
on our data is $H \approx 0.76$ (see Fig. \ref{fig8}).
Hence, the 
fluctuating rain rate alone produces an anomalous Hurst exponent, and 
the result obtained by Hurst is valid not only for the range of 1 year 
$< \tau <$ 1080 years that he considered but in fact also holds for 
$\tau$ = a few minutes to $\tau$ = $1/2$ year. Interestingly, the Hurst 
exponent deviates from this relation for $\tau < 10$ min, which is of 
the same order as the observed short-time trivial regime of the fractal 
dimension. 

To understand more precisely what is actually measured by the Hurst exponent, we applied the same 
method to a signal generated by swapping events and droughts at random. We kept the sizes 
and durations of rain events and droughts as determined from the real data and pasted them one after the other in random 
order. The Hurst exponent was \i{not} altered by this procedure. In this sense it is not a measure of correlation since it 
is 
not affected by the order in which events occur. In exactly what sense it measures persistence is part of our ongoing 
research. 
\section{Context} 
\label{context} 
Self-organised criticality offers the appropriate frame work for dealing with relaxational process with burst-like 
behaviour whose statistics are determined by scale invariant power laws \cite{18,19}. The term self-organised 
criticality 
refers to the tendency of many systems driven by an energy input at a slow and constant rate to enter states characterised 
by scale free behaviour. The statistics of the system then resemble those of a closed system near the critical point of a 
phase transition. 

A well-known example of such a process is the energy flow through the 
lithosphere, including the outermost crust of Earth. Tectonic plates are 
driven at a slow rate by currents in the asthenosphere, the liquid part 
below the lithosphere, which transports heat by currents. The energy 
transferred to the plates is intermediately stored in the form of 
tension until it is finally released in an earthquake. Earthquake 
statistics follow the Gutenberg-Richter power law that relates the 
seismic moment, a measure of the released energy to the probability of 
such an earthquake \cite{20}. Rather than a typical size with 
exponentially fewer larger than smaller quakes, scale invariant 
behaviour is observed. 

Given the right shape of grains, rice piles exhibit self-organised 
critical behaviour \cite{21}. Potential energy is added to the system by 
dropping rice grains onto the pile at a slow and constant rate. Due to 
the friction between individual grains, the pile builds up until its 
slope reaches a critical value. In this critical state, within the 
limits set by the system size, avalanches of all sizes are observed. 
During an avalanche, potential energy that was intermediately stored in 
the system is suddenly released in the form of heat. The distribution of 
energy release is once more a power law. 

Experiments on droplet avalanches show that self-organised criticality 
need not be restricted to granular media \cite{22}. Thresholds that 
enable the accumulation of energy before the release are given by 
surface tension and interface friction with other media.

Rain showers share many of the features of the above-mentioned systems (see Tab. \ref{tab1}). 
Two well separated time scales are present: The durations of drought 
periods, during which water evaporates, range up to months, while rain 
events take place on a much shorter time scale. The atmosphere receives 
a slow and constant energy input from the Sun's radiation. The absorbed 
energy evaporates water from the surface, which  is intermediately 
stored in the atmosphere. Note the analogy between liquid water 
in the atmosphere, tension in tectonic plates, and mass above the ground 
level in a granular pile. During a rain shower, the water mass that was 
slowly evaporated into the atmosphere, is suddenly released, and with it the original evaporation energy, i.e. the 
condensation energy. The power law observed for the size distribution of 
rain events is perfectly equivalent to the Gutenberg-Richter law in earthquake statistics. Just as plates of the 
lithosphere don't move smoothly along, there is no constant light rainfall balancing the evaporated water mass immediately 
at every moment in time. Rain events are relaxations in the sky.

\section{Conclusion} 
\label{conclusion} 
New insight into the working of rain can be gained by defining rain events, which can be regarded as energy relaxations 
similar to earthquakes or avalanches. Taking this perspective, scale-free power-law behaviour is found to govern the 
statistics of 
rain over a wide range of time- and event size scales. Where clear deviations from the observed power laws and fractal 
dimensions are found, the limits and peculiarities of the underlying dynamical system become apparent, and physical 
insight 
is gained. Rainfall time series cannot be reproduced by conventional methods of probability theory. To enable anything 
more 
than an explicit reproduction of the fractal properties, a deeper understanding of self-organising processes leading to  
fractality must be sought. Our findings suggest that rain is an excellent example of a self-organised critical process. 
Rain is a  ubiquitous phenomenon, and data collection is relatively easy. It is therefore well suited for work on 
self-organised criticality. For our purposes, the remote sensing technique employed by the MRR-2 has proved extremely 
powerful. The radar is capable of even higher temporal resolution than $1$ min, limited only by the finite height of the 
scattering volume, and achieves outstanding precision in the low-intensity limit. Comparison with data from other 
measuring 
sites, especially from warmer regions without snow and regions with more periodic climate would be useful in order to 
answer questions regarding 
the universality of the observed features.

\section{Acknowledgments} 
\label{acknowledgments} 
O.P. would like to thank the group in Hamburg and especially G. Peters for their contributions to the success of the 
project. K.C. gratefully acknowledges the financial support of U.K. EPSRC through grant nos. GR/R44683/01 and 
GR/L95267/01. 
The MRR-2 data collection was supported by the EU under the Precipitation and Evaporation Project (PEP) in BALTEX. 

$^*${To whom correspondence should be addressed, 
E-mail: ole.peters@ic.ac.uk}

\widetext
\begin{table} 
\caption{Rain events are analogous to a variety of relaxational 
processes in nature. The two best known examples of such processes, earthquakes and avalanches in granular media, are 
summarised above.}
	\begin{tabular}{cccc}
 		System&Crust of Earth& Granular Pile &Atmosphere\\ \hline
 		Energy Source & Currents in asthenosphere & Addition of grains & Sun  \\
 		Energy Storage & Tension & Gravitational potential & Evaporated water\\ 
 		Threshold & Friction & Friction & Saturation \\ 
 		Release of Energy & Earthquake & Avalanche & Rain event\\ 
 	\end{tabular}
\label{tab1}
\end{table}
 

\begin{thebibliography}{40} 

\bibitem{1} 
H.E. Hurst, 
{\em Long-Term Storage: An Experimental Study} 
(Constable \& Co. Ltd., London, 1965). 

\bibitem{2}
O. Peters, C. Hertlein, and K.Christensen, 
Phys. Rev. Lett. {\bf 88}, 018701-1 (2002).

\bibitem{3} 
Global Terrestrial Observation System: Requirements for precipitation 
measurements  http://www.fao.org/gtos/tems/variable\_list.jsp (2001). 

\bibitem{4} 
D. Klugmann, K. Heinsohn, and H.J. Kirtzel, 
Contr. Atmos. Phys. {\bf 69}, 247 (1996). 

\bibitem{5} 
MMR-2, Physical Basis, pp. 21 (1998). 
Available from METEK GmbH, Fritz-Stra{\ss}mann-Stra{\ss}e 4, D-25337 
Elmshorn, Germany. 

\bibitem{6} 
Information on the BALTEX project is available from the website 
http://w3.gkss.de/baltex/. 

\bibitem{7} 
D. Atlas, R.C. Srivastava, and R.S. Sekhon, 
Rev. Geophys. Space Phys. {\bf 11}, 1 (1973). 

\bibitem{8} 
G. Mie, Ann. Phys. {\bf 25}, 377 (1908).

\bibitem{9}
K. Beard and C. Chuang,
J. Atmos. Sci. {\bf 44} (11) 1509 (1987).

\bibitem{10}
K. Beard, 
J. Atmos. Oceanic Technol. {\bf 2}, 468 (1985).

\bibitem{11}
R.C. Weast, {\em CRC Handbook of Chemistry and Physics 58th edition}, (CRC Press Cleveland, 1978).

\bibitem{12} 
F. Schmitt, S. Vannitsem, and A. Barbosa, 
J. Geophys. Res. {\bf 103}, 23,181, 1998. 

\bibitem{13}
R.A. Houze, Jr. {\em Cloud Dynamics} (Academic Press Inc, San Diego, 1993).

\bibitem{14} 
J. Feder, {\em Fractals} (Plenum Press, New York, 1988). 

\bibitem{15} 
E.V. Laursen, J. Larsen, K. Rajakumar, J. Cappelen and T. Schmith, 
Observed Daily Precipitation, Temperature and Cloud Cover for Seven 
Danish Sites, 1876-2000. 
Technical report from DMI. Data available from http://www.dmi.dk/dmi via 
"publikationer", "tekniske rapporter", No. 01-10. 

\bibitem{16}
E.X. Berry and R.L. Reinhardt, 
Modelling of condesation and collection within clouds. Final report, NSF grant GA-21350 (1973).

\bibitem{17} 
W Feller, Ann. Math. Stat. {\bf 22}, 427 (1951). 

\bibitem{18} 
H.J. Jensen, {\em Self-Organized Criticality: Emergent Complex 
Behavior in Physical and Biological Systems} (Cambridge University 
Press, 1998). 

\bibitem{19}
P. Bak, {\em How nature works: The science of self-organised criticality} 
(Springer Verlag New York, 1996).

\bibitem{20} 
B. Gutenberg and C.F. Richter, 
Bull. Seismol. Soc. Am. {\bf 34}, 185 (1994). 

\bibitem{21} 
V. Frette, K. Christensen, A. Malthe-S{\o}renssen, J. Feder, T. 
J{\o}ssang, 
and P. Meakin, 
Nature {\bf 379}, 49 (1996). 

\bibitem{22} 
B. Plourde, F. Nori, and M. Bretz, Phys. Rev. Lett {\bf 71}, 2749 (1993). 
\end{thebibliography}
\end{document}